\documentstyle[11pt,newpasp,twoside]{article}
\def\edcomment#1{\iffalse\marginpar{\raggedright\sl#1\/}\else\relax\fi}
\reversemarginpar

\begin{document}
\title{Ground-based exoplanet near-infrared search by imaging and spectroscopy:
3 new companion candidates in TWA}
\author{Ralph Neuh\"auser}
\affil{MPI extraterrestrische Physik, D-85741 Garching, Germany, 
and University of Hawaii, Institute for Astronomy, Honolulu, USA}
\author{Nuria Hu\'elamo}
\affil{MPI extraterrestrische Physik, D-85741 Garching, Germany}
\author{Eike W. Guenther}
\affil{Th\"uringer Landessternwarte Tautenburg, D-07778 Tautenburg, Germany}
\author{Wolfgang Brandner, Jo\~ao Alves, Fernando Comer\'on}
\affil{European Southern Observatory, D-85748 Garching, Germany}
\author{Monika G. Petr}
\affil{MPI f\"ur Radioastronomie, Auf dem H\"ugel 69, D-53121 Bonn, Germany}
\author{Jean-Gabriel Cuby}
\affil{European Southern Observatory, Santiago, Chile}

\begin{abstract}
We report first results from our ground-based infrared imaging search for 
sub-stellar companions (brown dwarfs and giant planets) of young (up to 
100 Myrs) nearby (up to 100 pc) stars, where companions should be well 
separated from the central stars and still relatively bright due to ongoing 
accretion and/or contraction. Our observations are performed mainly 
with SOFI and SHARP at the ESO 3.5m NTT on La Silla (imaging) and with 
ISAAC at the ESO 8.2m Antu (VLT-UT1) on Cerro Paranal (imaging and 
spectroscopy), all in the H- and K-bands.
Here, we present new companion candidates\footnote{Based on observations obtained 
at the European Southern Observatory on Cerro Paranal and La Silla in program 65.L-0144.}
around three T Tauri stars (TWA-8~A, RXJ1121.1-3845, and RXJ1121.3-3447~N)
in the TW Hya group, which would have been sub-stellar if at the same distance 
and age as the T Tauri stars, but are found to be background stars by spectroscopy.
\end{abstract}

\section{Introduction}

Sub-stellar companions like brown dwarfs and giant planets are hard to find 
due to the problem of dynamic range: Faint objects very close to much brighter stars.
All very nearby stars, within $\sim 25~pc$, are $\ge 1$ Gyr old.
One way around this problem is to search for sub-stellar companion candidates around
very young relatively nearby stars.
There are indeed young nearby stars, like GJ 182 ($\sim 20$ Myrs at 27 pc),
the Tucanae and HorA groups of young zero-age main-sequence stars,
the TW Hya association (TWA) of about one dozen pre-main sequence stars 
isolated from cloud material,
and the MBM 12 cloud at $\sim 65~pc$ with several classical and weak-line T Tauri stars.

We have started a direct imaging search for sub-stellar companions around those
young nearby stars: Up to 100 Myrs and up to 100 pc.
Including new nearby stars discovered recently among ROSAT X-ray sources,
there are more than 100 such stars known to date.
We use the infrared imaging camera SofI at the ESO 3.5m NTT on La Silla,
the MPE speckle camera SHARP (also used at NTT),
the infrared imager and spectrograph ISAAC at the ESO 8.2m Antu (VLT-UT1)
on Cerro Paranal, and the adaptive optics infrared instrument ALFA
on the Calar Alto 3.5m telescope.
For companions candidates found in the 1st epoch imaging, we then take a
2nd epoch image one or a few years later for proper motion confirmation
and/or an infrared spectrum to confirm or reject their sub-stellar nature.

Five sub-stellar companions have been confirmed so far by both spectroscopy
and proper motion: After Gl~229~B,
G~196-3~B,
and Gl~570~D,
the two youngest brown dwarf companions known so far were confirmed by us
using spectroscopy and proper motion, namely
CoD$-33 ^{\circ} 7795$~B in TWA (Neuh\"auser et al. 
2000b)\footnote{Schneider et al. (2001) 
confirmed the M8.5-M9 spectral type reported by Neuh\"auser et al. (2001) with
an HST/STIS spectrum with a smaller wavelength range.} 
and HR~7329~B in Tucanae (Guenther et al. 2001).
A giant planet candidate near TWA-7 (Neuh\"auser et al. 2000a) has been
found to be a background K-type star in an H-band spectrum taken with
ISAAC at the VLT (Neuh\"auser et al. 2001).

Here, we report on three more sub-stellar companion candidates found by direct imaging
near three other TWA members.

\section{More companion candidates in TWA}

Using the infrared imager Son of Isaac (SofI) at the ESO 3.5m New
Technology Telescope (NTT) on La Silla we took images of several TWA members.
The observations were made on 2000 May 17 and 19 with 1.0" to 1.1" seeing.
We used the H-band filter and the small SofI field (0.147"/pix)
with a total exposure time of 10 min per target ($460 \times 1.3$ sec in auto-jitter mode).
Darks, flats, and standard stars were taken in the same night, and we performed 
standard data reduction with {\em eclipse}\footnote{www.eso.org/eclipse} 
and MIDAS. We found new companion candidates 
around TWA-8 (a TWA T Tauri star found by Webb et al. 1999),
RXJ1121.1-3845 (=GSC 7739 2190), and RXJ1121.3-3447 (=GSC 7210 1352), 
originally found as new T Tauri stars by W. Hoff (PhD thesis) 
and published by Sterzik et al. (1999), data are listed in table 1.

\begin{table} 
\begin{tabular}{lcc|ccc|c} 
\multicolumn{7}{c}{\bf Table 1: Three new companion candidates (=~c/c) in TWA} \\ \hline
\multicolumn{3}{c}{Primary star} &  \multicolumn{3}{c}{separation (*)}         & c/c \\
Name & Sp.Type & H [mag]         & $\Delta \alpha$ & $\Delta \delta$ & sep ["] & H [mag] \\ \hline
TWA-8 & M2,M5 & 7.8,9.4 & 5.6" E & 8.3" N & 10.0 & 15.5 \\ 
RXJ1121.1-3845 & M2 & 8.4 & 6.0" W & 2.6" S & 6.5 & 16.5 \\
RXJ1121.3-3447 & M1,M2 & 7.7,7.8 & 5.4" W & 9.1" N & 10.6 & 16.8 \\ \hline
\end{tabular}
{\small Remarks: (*) TWA-8 and RXJ1121.2-3446 are binary stars, the separations given are 
measured from the c/c to the primary (northern) star. All mags $\pm 0.1$ mag, sep $\pm 0.1$".
The companion candidate to RXJ1121.1-3845 was found also by D. Trilling \& R. 
Jayawardhana using the cold coronograph CoCo at the NASA IRTF (priv. com.).}
\end{table}

If the companion candidates would be bound, i.e. at $\sim 55~pc$ and $\sim 10$ Myrs as TWA,
their magnitudes would correspond to objects with only a few Jupiter masses, according to 
different sub-stellar tracks and isochrones, i.e. they would be sub-stellar. 
We took follow-up H-band spectra using ISAAC at Antu (VLT-UT1)
and found that all three companion candidates are unrelated background stars
(to be reported in more detail later).
No companion candidates were found within 15" around RXJ1109.7-3907
(PPM 288568, H$=8.8$ mag), the other new TWA T Tauri found by Hoff and Sterzik et al.,
down to $H \simeq 17$ mag.

\section{Summary}

Our observations of TWA-7 and TWA-5 (as well as those presented here),
all performed with ground-based present-day technology, show that objects
as faint as expected for young nearby giant planets located very close (few arc sec) 
to much brighter stars ($\Delta$ mag $\simeq 10$ mag) can be detected already.
Once such a planetary companion candidate is confirmed,
we can probe its atmosphere by spectroscopy.

\end{document}